\documentclass[oribibl,fleqn]{llncs}
\usepackage[utf8]{inputenc}
\usepackage{ifthen}
\usepackage{graphicx}
\usepackage{isabelle,isabellesym}
\usepackage{pdfsetup}
\isabellestyle{literal}

\isadroptag{theory}

\renewcommand{\isastyletxt}{\isastyletext}
\renewcommand{\isadigit}[1]{\isamath{#1}}

\newcommand{\secref}[1]{\S\ref{#1}}
\newcommand{\figref}[1]{fig.~\ref{#1}}

\newcommand{\BibTeX}{Bib{\TeX}}

\newcommand{\para}[1]{\medskip\par\noindent\textbf{#1}}

\newcommand{\success}{\para{Success:}}
\newcommand{\failure}{\para{Failure:}}
\newcommand{\changes}{\para{Changes:}}
\newcommand{\future}{\para{Future:}}

\begin{document}

\title{Interaction with Formal Mathematical Documents in Isabelle/PIDE}
\author{Makarius Wenzel}
\institute{\url{https://sketis.net}}
\maketitle

\begin{abstract}

Isabelle/PIDE has emerged over more than 10 years as the standard Prover IDE
for interactive theorem proving in Isabelle. The well-established Archive of
Formal Proofs (AFP) testifies the success of such applications of formalized
mathematics in Isabelle/HOL. More recently, the scope of PIDE has widened
towards languages that are not connected to logic and proof in Isabelle, but
taken from a broader repertoire of mathematics on the computer. The present
paper provides a general overview of the PIDE project and its underlying
document model, with built-in parallel evaluation and asynchronous
interaction. There is also some discussion of original aims and approaches,
successes and failures, later changes to the plan, and ideas for the future.

\end{abstract}

\begin{isabellebody}%
\setisabellecontext{Paper}%
\isadelimtheory
\isanewline
\isanewline
\endisadelimtheory
\isatagtheory
\isacommand{theory}\isamarkupfalse%
\ Paper\isanewline
\ \ \isakeyword{imports}\ Main\isanewline
\isakeyword{begin}%
\endisatagtheory
{\isafoldtheory}%
\isadelimtheory
\endisadelimtheory
\isadelimdocument
\endisadelimdocument
\isatagdocument
\isamarkupsection{Introduction%
}
\isamarkuptrue%
\endisatagdocument
{\isafolddocument}%
\isadelimdocument
\endisadelimdocument
\begin{isamarkuptext}%
Isabelle/PIDE means \textbf{Prover IDE}: its implementation relies on
  Isabelle/Scala, and the standard front-end is Isabelle/jEdit: so all these
  brand names can be used interchangeably at some level of abstraction. The
  presentation at Schloss Dagstuhl in October 2009 \cite{Wenzel:2009:Dagstuhl} provides an interesting historical view of the
  initial concepts of Isabelle/Scala and the preliminary implementation of
  Isabelle/jEdit. Work on that had already started one year earlier, so the
  Dagstuhl presentation in August 2018 could use the title ``The Isabelle
  Prover IDE after 10 years of development'' \cite{Wenzel:2018:Dagstuhl}.
  In the years between, there have been many papers about the project, notably
  \cite{Wenzel:2010,Wenzel:2011:CICM,Wenzel:2012:CICM,Wenzel:2012:UITP-EPTCS,Paral-ITP:CICM2013,Wenzel:2014:UITP,Wenzel:2018:FIDE,Wenzel:2018:UITP}.

  Considerable complexity of Isabelle/PIDE concepts and implementations has
  accumulated over time, and presenting a comprehensive overview in this paper
  poses a challenge. Subsequently, we start with two concrete application
  scenarios: standard Isabelle/jEdit (\secref{sec:isabelle-jedit}) and
  non-standard Isabelle/Naproche (\secref{sec:naproche}). More systematic
  explanations of the PIDE document-model are given in \secref{sec:PIDE}.
  Discussion of aims and approaches of PIDE follows in
  \secref{sec:approaches}: this provides a perspective on design decisions
  from the past, with projections into the future.%
\end{isamarkuptext}\isamarkuptrue%
\isadelimdocument
\endisadelimdocument
\isatagdocument
\isamarkupsubsection{Isabelle/PIDE as IDE for Interactive Proof Documents \label{sec:isabelle-jedit}%
}
\isamarkuptrue%
\endisatagdocument
{\isafolddocument}%
\isadelimdocument
\endisadelimdocument
\begin{isamarkuptext}%
Isabelle is an interactive proof assistant (similar to Coq or HOL4), and
  PIDE is the Prover IDE framework for it. Isabelle/PIDE is implemented in
  Isabelle/ML (based on Poly/ML) and Isabelle/Scala (on the Java Virtual
  Machine). This arrangement allows to use existing IDE front-ends from the
  Java ecosystem, e.g. the plain text editor jEdit (\url{http://jedit.org}). The
  combined Isabelle/jEdit \cite{isabelle-jedit} is presently the most
  sophisticated application of PIDE, and the default user-interface for
  Isabelle. There are other PIDE front-ends, e.g. Isabelle/VSCode and a
  headless server, but non-PIDE interaction has already been discontinued in
  October 2014. Consequently, the classic Proof~General Emacs \cite{Aspinall:TACAS:2000} does not work for Isabelle anymore: it was based on
  the TTY-loop that no longer exists.

  \begin{figure}[htb]
  \centering
  \includegraphics[width=\textwidth]{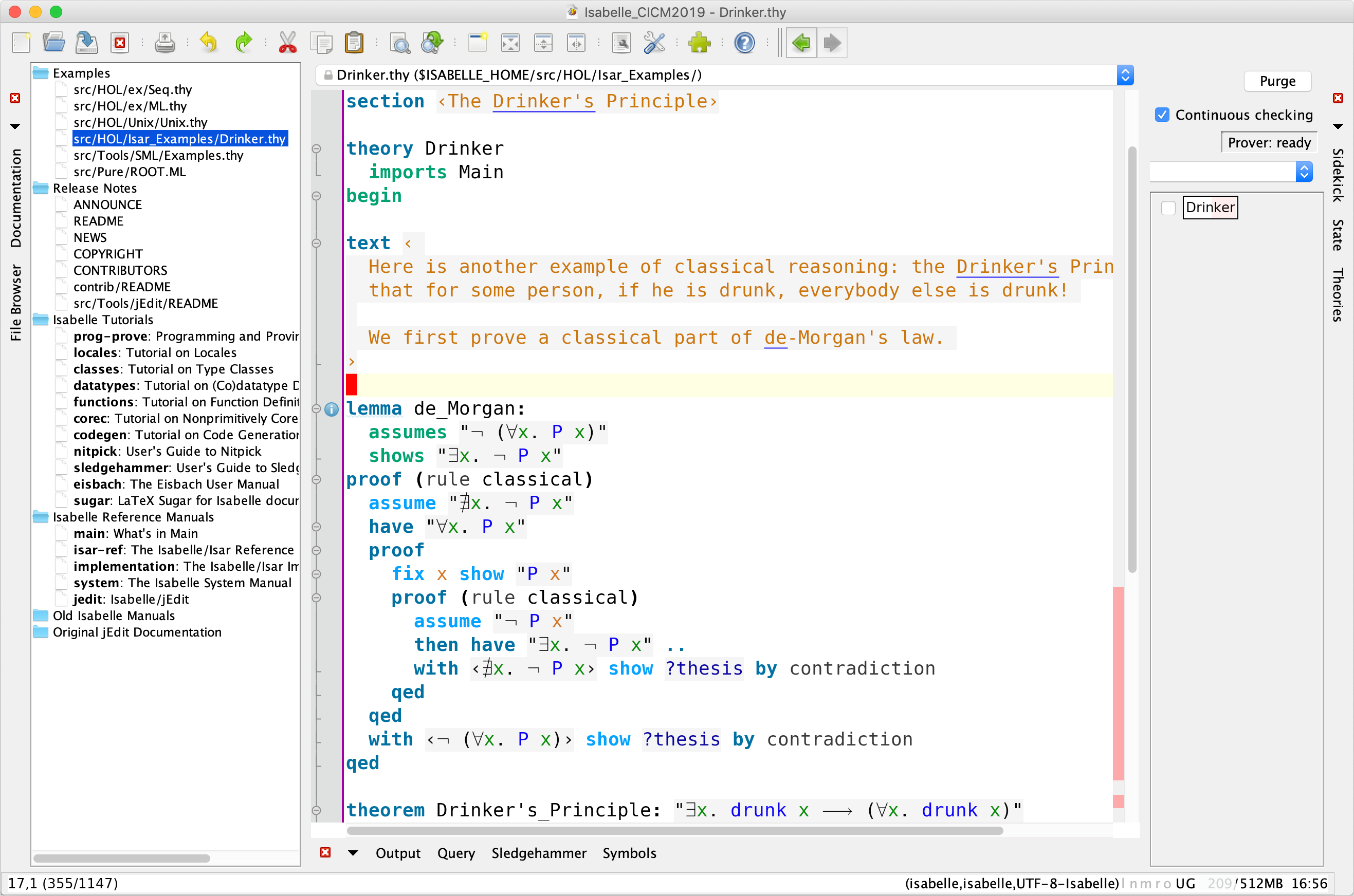}
  \caption{The all-inclusive Isabelle/PIDE/jEdit application}
  \label{fig:isabelle-jedit}
  \end{figure}

  Users who download\footnote{\url{https://isabelle.sketis.net/Isabelle_CICM2019}} and
  run the application bundle for their respective operating-system first
  encounter the Isabelle/jEdit desktop application, similar to
  \figref{fig:isabelle-jedit}: it provides immediate access to documentation,
  examples, and libraries. All of these are \emph{Formal Mathematical
  Documents} in Isabelle, which are organized as \emph{theories} and \emph{sessions}
  (i.e. collections of theories with optional document output). This includes
  the Isabelle manuals from the \isatt{Documentation} panel, which shows PDFs
  generated from sessions in the \isatt{{\char`\$}ISABELLE{\char`\_}HOME/\discretionary{}{}{}src/\discretionary{}{}{}Doc} directory.

  By opening \isatt{{\char`\$}ISABELLE{\char`\_}HOME/\discretionary{}{}{}src/\discretionary{}{}{}Doc/\discretionary{}{}{}JEdit/\discretionary{}{}{}JEdit.thy} in Isabelle/jEdit, we
  see the theory sources of the Isabelle/jEdit manual in the Prover IDE. Its
  content mainly consists of traditional document structure: section headings
  and blocks of text. Such \emph{quotations} of informal text may also contain
  formal items via \emph{antiquotations}. The latter concept was introduced to
  allow prose text to talk about logical terms and types, but the same
  mechanism is re-used to augment {\LaTeX} by formal elements: links to files
  or URLs, text styles with robust nesting (bold, emphasized, verbatim,
  footnote), item lists as in Markdown, citation management wrt. {\BibTeX}
  databases etc. (see also \cite[Chapter 4]{isabelle-jedit}).

  Beyond self-application of Isabelle/PIDE/jEdit to its own documentation, the
  Isabelle distribution provides libraries and applications of formalized
  mathematics, mostly in Isabelle/HOL (see the directory
  \isatt{{\char`\$}ISABELLE{\char`\_}HOME/\discretionary{}{}{}src/\discretionary{}{}{}HOL}). The material may be edited directly in the
  Isabelle/jEdit Prover IDE --- except for the HOL session itself, which is
  preloaded as non-editable session image. Such spontaneous checking may
  require substantial hardware resources, though. E.g.
  \isatt{{\char`\$}ISABELLE{\char`\_}HOME/\discretionary{}{}{}src/\discretionary{}{}{}HOL/\discretionary{}{}{}Analysis/\discretionary{}{}{}Analysis.thy} works best with 8 CPU cores
  and 16\,GB main memory, and still requires several minutes to complete. Note
  that this is not just browsing, but \emph{semantic editing} of a live document:
  a checked state provides full access to the execution environment of the
  prover.

  Development of complex proof documents requires add-on tools: a theory
  library usually provides new logical content together with tools for
  specifications and proofs. Isabelle/HOL itself is an example for that, with
  many \emph{proof methods} to support (semi-)automated reasoning in Isabelle/Isar
  \cite{Wenzel:2006:Festschrift}, and \emph{external provers} (ATP, SMT) for
  use with \emph{Sledgehammer} \cite{isabelle-sledgehammer}. Isabelle/PIDE
  orchestrates all tools within one a run-time environment of parallel
  functional programming. Results are exposed to the front-end via a stream of
  \emph{PIDE protocol} messages. The editor can retrieve the resulting PIDE
  document markup in real-time (without waiting for the prover) and use
  conventional GUI elements to show it to the user: e.g. as text colours,
  squiggly underlines, icons, tooltips, popups. Output generated by the prover
  can have extra markup to make it \emph{active}: when the user clicks on it,
  edits will be applied to the text to continue its development, e.g. see
  \cite[\S3.9]{isabelle-jedit} for document-oriented interaction with
  Sledgehammer.

  \medskip The example sessions of the Isabelle distribution are quite substantial,
  but most Isabelle/HOL formalizations are now maintained in \textbf{AFP}, the
  \emph{Archive of Formal Proofs} \cite{AFP}. AFP is organized like a scientific
  journal, not a repository of ``code''. Thus it is similar to the Mizar
  Mathematical Library\footnote{\url{http://mizar.org/library} and
  \url{http://mizar.org/fm}}, but with more flexibility and programmability of
  \emph{domain-specific formal languages} \cite{Wenzel:2018:FIDE}. That
  continues the original LCF/ML approach \cite{Gordon-Milner-Wadsworth:1979,Gordon-Milner-etal:1978} towards active documents with full-scale
  Prover IDE support.

  AFP version 28e97a6e4921 (April 2019) has 315 authors, 471 sessions, 4912
  theories. In principle, it is possible to load everything into a single
  prover session for Isabelle/jEdit. But scaling is not for free, and doing
  that blindly requires two orders of magnitude more resources than for
  HOL-Analysis above. In practical development of large AFP entries, users
  still need some planning and manual arrangement of sessions, to restrict the
  focus to relevant parts of AFP.

  A truly integrated development environment should do that automatically for
  the user, and treat Isabelle + AFP as one big mathematical document for
  editing (and browsing). Concrete ideas for further scaling of the PIDE
  technology are outlined in \cite{Wenzel:2018:Isabelle}, but it will
  require some years to get there.%
\end{isamarkuptext}\isamarkuptrue%
\isadelimdocument
\endisadelimdocument
\isatagdocument
\isamarkupsubsection{Isabelle/Naproche for Automatic Proof-Checking of Ordinary
  Mathematical Texts \label{sec:naproche}%
}
\isamarkuptrue%
\endisatagdocument
{\isafolddocument}%
\isadelimdocument
\endisadelimdocument
\begin{isamarkuptext}%
Naproche-SAD is a recent tool by Frerix and Koepke \cite{Frerix-Koepke:2018}, based on the original \emph{System for Automated
  Deduction} (SAD) by Paskevich and others \cite{Paskevich-et-al:2004}. It
processes the \emph{Formal Theory Language} (ForTheL), which is designed to look
like mathematical text, but it is restricted to a small subset of natural
language.

The tool is implemented in Haskell as a plain function from input text to
output messages. A file is like a chapter of mathematical text, with a
nested tree-structure of elements and sub-elements (for signatures,
axiomatizations, statements, proofs). Output messages inform about the
translation of mathematical text to problems of first-order logic, and
indicate success or failure of external proof checking; the latter is
delegated to the \emph{E Prover} by Stephan Schulz and can take several seconds
for each proof obligation.

To integrate Naproche-SAD into PIDE, Frerix and Wenzel have reworked the
Haskell program over 2 months in 2018, to turn the command-line tool into a
service for reactive checking of ForTheL texts. Isabelle integration was
done via the new Isabelle/Haskell library and some glue code in
Isabelle/Scala to register ForTheL as auxiliary file-format (extension
\isatt{.ftl}).

\begin{figure}[htb]
\centering
\includegraphics[width=\textwidth]{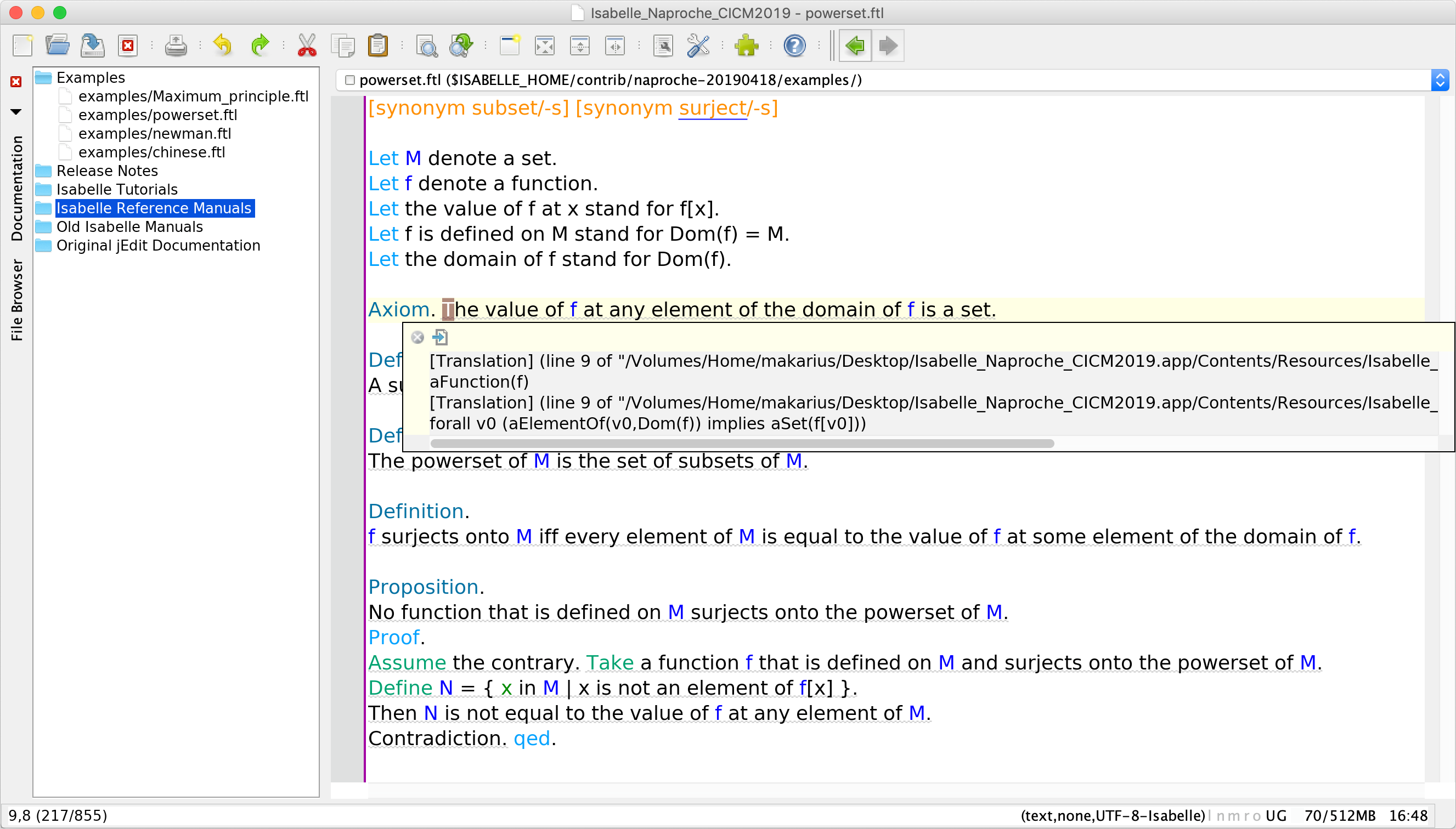}
\caption{Isabelle/Naproche with ``ordinary mathematical text''}
\label{fig:isabelle-naproche}
\end{figure}

The resulting Isabelle/Naproche application is available as multi-platform
download\footnote{\url{https://isabelle.sketis.net/Isabelle_Naproche_CICM2019}}. A
running instance is shown in \figref{fig:isabelle-naproche}: users can
directly open ForTheL files (e.g. from \isatt{Documentation} / \isatt{Examples}) and
wait briefly to see output messages attached to the text in the usual IDE
manner. Further edits produce a new version of the text, which is sent in
total to Naproche-SAD again. The back-end is sufficiently smart to avoid
redundant checking of unchanged sub-elements: it keeps a global state with
results of old versions: this is easy to implement as the program keeps
running until shutdown of Isabelle/PIDE.

The general approach of using an external tool in Isabelle/PIDE like a
function on input sources to output messages has been demonstrated before
for \isatt{bibtex} databases \cite[\S5]{Wenzel:2018:FIDE}. This technique
depends on the following conditions:

\begin{itemize}%
\item Input and output works via temporary files or anonymous streams for each
invocation --- no global state within the file-system.

\item Source positions in output uses precise text offsets --- not just line
numbers: this typically requires careful inspection of text encoding and
line-endings.

\item The tool starts quickly and runs only briefly. As a rule of thumb for
IDE reactivity, 10\,ms is very fast, 100\,ms is fast enough, 500\,ms is
getting slow.

\item A long-running tool needs to be interruptible via POSIX signals: C/C++
programs often require some corrections in this respect.

\item A long-running tool should output messages incrementally on a stream to
let the user see approximative PIDE markup early. E.g. syntax markup
immediately after parsing, and the status of semantic checking later on.%
\end{itemize}

The small and quick \isatt{bibtex} program can be started afresh for each version
of input text, and messages returned as a single batch extracted from the
log file. In contrast, the rather heavy \isatt{Naproche{\char`\-}SAD} executable can spend
several seconds on small mathematical texts, so the continuously running
server process with its internal cache is important for reactivity; it also
avoids repeated startup of a big program. To achieve that, the Haskell
program had to be changed significantly, but typed functional programming
helps to keep an overview of global state and interactions with the world:
this makes it easy to isolate the main function for concurrent invocations
of multiple server requests.

In summary, we see that Isabelle is not just Isabelle/HOL: Isabelle/Naproche
uses logic in its own way, independently of the Isabelle logical framework.%
\end{isamarkuptext}\isamarkuptrue%
\isadelimdocument
\endisadelimdocument
\isatagdocument
\isamarkupsection{The PIDE Document Model \label{sec:PIDE}%
}
\isamarkuptrue%
\endisatagdocument
{\isafolddocument}%
\isadelimdocument
\endisadelimdocument
\begin{isamarkuptext}%
Abstractly, we can understand a PIDE document as large expression of
  \emph{embedded sub-languages} that can be \emph{explored interactively} in the
  editor, while the prover is processing the \emph{execution} as a parallel
  functional program. Continued editing by the user may cancel ongoing
  executions by the prover, and replace no longer relevant parts of the
  document by new material to be checked eventually. The communication of the
  editor with the prover always works via \emph{document edits}, e.g. to insert or
  remove segments of \emph{text} to produce a new \emph{document version}. Edits also
  determine the \emph{perspective} on the text: it tells the prover which parts of
  the document are important for the user. E.g. scrolling an editor window
  will update the document and refocus the execution to include uncovered
  text.

  Document content is represented as \emph{plain text} within editor buffers, but
  it is also possible to include \emph{auxiliary files} that could be external
  blobs in some user-defined format (usually they are plain text as well).
  Both input sources and output messages are rendered in enhanced editor
  buffers to support text decorations and pretty-printing in the style of
  Oppen \cite{Oppen:1980}: its formatting and line-breaking dynamically
  adjusts itself to font metrics and window sizes.

  PIDE markup (XML) over input sources or within output pretty-trees is used
  for text colours, tooltips, popups etc. Thus the syntax-highlighting in
  jEdit is augmented by \emph{semantics-rendering} in Isabelle/jEdit.%
\end{isamarkuptext}\isamarkuptrue%
\isadelimdocument
\endisadelimdocument
\isatagdocument
\isamarkupsubsection{Document Structure and Organization: Theories and Sessions%
}
\isamarkuptrue%
\endisatagdocument
{\isafolddocument}%
\isadelimdocument
\endisadelimdocument
\begin{isamarkuptext}%
PIDE documents follow the traditional structure of mathematical texts, which
  is a sequence of ``definition--statement--proof'' given in foundational
  order.

  A \emph{definition} is an atomic text element that extends the theory context:
  this could be a genuine definition of a constant in Isabelle/HOL, or the
  definition of a module in Isabelle/ML, or something else in a
  domain-specific formal language. A \emph{statement} is a pending claim within
  the theory that requires justification by the subsequent \emph{proof}: it could
  be a theorem statement in the logic or a derived definitional specification
  (e.g. a recursive function with termination proof). Definitions and
  statements can be arbitrary user-defined notions of the Isabelle
  application, but proofs always use the Isabelle/Isar proof language \cite{Wenzel:2006:Festschrift}.

  \emph{Foundational order} means that the overall document elements can be seen
  as a well-founded sequence of elementary constructions: mutual recursion is
  restricted to a single definitional element, e.g. one that introduces
  multiple recursive functions. For better scalability, projects are usually
  built up as an \emph{acyclic directed graph} of theory nodes: each theory
  imports existing theories and forms a canonical \emph{merge} of the contexts
  before adding further material. Isabelle context management has a built-in
  notion of monotonicity to propagate results: after import, a well-formed
  term remains well-formed, a proven theorem remains proven etc. Entities from
  domain-specific languages need to conform to the same abstract principle of
  ``monotonic reasoning''.

  \smallskip A \emph{session} consists of a sub-graph of imported theories, together with
  auxiliary files for document-preparation in {\LaTeX}. This roughly
  corresponds to a ``project'' in conventional IDEs for programming languages.
  A \emph{session image} is a ``dumped world'' of the underlying Isabelle/ML
  process of a session, to speed-up reloading of its state. Each session image
  may refer to a single parent, to form an overall \emph{tree} of session images.
  Merging session images is \emph{not} supported, but a session can have multiple
  \emph{session imports} to load theories from other sessions. This is not a copy
  of the other theory, but an inclusion that requires re-checking for the new
  session (often this is faster than building a complex stack of session
  images).

  In AFP, each entry usually consists of a single session for its formal
  theory content, together with the {\LaTeX} setup for presentation as a
  journal article \cite{AFP}.

  \smallskip In addition to session documents, Isabelle2019 supports \emph{session
  exports}: tools in Isabelle/ML may publish arbitrary blobs for the \emph{session
  database} (with optional XZ compression). For example, the AFP entry
  \isatt{Buchi{\char`\_}Complementation} generates a functional program from its
  specification and exports the compiled executable. The session export
  mechanism avoids uncontrolled write-access to the global file-system: since
  PIDE document processing operates concurrently on multiple versions, writing
  out physical files would be ill-defined.

  In \emph{batch mode} (e.g. \isatt{isabelle\ build}), session sources are read from the
  file-system, and session exports are added to the SQLite database for this
  session build. Command-line tools like \isatt{isabelle\ export} or
  \isatt{isabelle\ build\ {\char`\-}e} can retrieve exports later on: here the session
  database acts like a \isatt{zip} archive.

  In \emph{interactive mode} (e.g. \isatt{isabelle\ jedit}), session sources are managed
  by the editor (with backing by the file-system), and session exports are
  stored in Isabelle/Scala data structures. Isabelle/jEdit provides virtual
  file-systems with URL prefixes \isatt{isabelle{\char`\-}session:} and \isatt{isabelle{\char`\-}export:}
  to access this information interactively, e.g. see the jEdit
  \isatt{File\ Browser}, menu \isatt{Favorites}, the last two items. An example is
  session \isatt{Tools/Haskell}, theory \isatt{Haskell}, command
  \isa{\isacommand{export{\isacharunderscore}generated{\isacharunderscore}files}} at the bottom: it exports generated sources of the
  Isabelle/Haskell library for re-use in other projects, e.g. Naproche-SAD
  (\secref{sec:naproche}).%
\end{isamarkuptext}\isamarkuptrue%
\isadelimdocument
\endisadelimdocument
\isatagdocument
\isamarkupsubsection{Common Syntax for Embedded Languages: Cartouches \label{sec:cartouches}%
}
\isamarkuptrue%
\endisatagdocument
{\isafolddocument}%
\isadelimdocument
\endisadelimdocument
\begin{isamarkuptext}%
The \emph{outer syntax} of Isabelle theories is the starting point for
  user-defined language elements: the header syntax \isa{\isacommand{theory}\ A\ \isakeyword{imports}\ B\isactrlsub {\isadigit{1}}\ {\isasymdots}\ B\isactrlsub n} is hardwired, but everything else is a defined \emph{command} within the
  theory body. The Isabelle bootstrap provides the initial \isa{\isacommand{ML}} command:
  Isabelle/Pure and Isabelle/HOL are using that to define a rich collection of
  commands that users often understand as \emph{the} Isabelle theory language, but
  it is merely a library.

  A command definition requires a \isa{\isakeyword{keywords}} declaration in the theory
  header, and a \emph{command parser} with semantic \emph{command transaction} in the
  theory body. All command parsers operate on the token language of
  Isabelle/Isar: it provides identifiers, numerals, quoted strings, embedded
  source etc.

  \emph{Quoted string} tokens are similar to string literals in ML. Some decades
  ago, there was no outer syntax and everything embedded into ML like that.
  Today we still see embedded types, terms, and propositions in that historic
  notation. Nested quotations do work, but require awkward backslash-escapes
  for quotes: the number of backslashes is exponential in the depth of
  nesting, so only one or two levels are seen in practice.

  \emph{Embedded source} tokens use the \emph{cartouche} notation of Isabelle, which
  was introduced a few years ago to facilitate arbitrary nesting. The quotes
  are directed (like open/close parentheses) and chosen carefully to remain
  outside of usual application languages. The Isabelle symbols \isatt{{\char`\\}}\isatt{{\char`\<}open{\char`\>}}
  and \isatt{{\char`\\}}\isatt{{\char`\<}close{\char`\>}} are used for that; they are rendered in the front-end as
  French single-quotes: e.g. \isa{{\isasymopen}source{\isasymclose}}. Thus a command that takes a cartouche
  as outer syntax remains free to use its own lexical conventions in the
  nested source.

  For example, the subsequent ML snippet defines a term \isa{t} within the ML
  environment of the theory; the \isa{\isactrlterm }-antiquotation inside ML uses
  regular term notation to construct the corresponding ML datatype value;
  there are further nested cartouches for comments inside these
  domain-specific languages.%
\end{isamarkuptext}\isamarkuptrue%
\isadelimML
\ \ %
\endisadelimML
\isatagML
\isacommand{ML}\isamarkupfalse%
\ {\isacartoucheopen}val\ t\ {\isacharequal}\ \isactrlterm {\isasymopen}{\isasymlambda}x{\isachardot}\ x\ {\isasymle}\ y\ {\isacharplus}\ z\ %
\isamarkupcmt{comment in term%
}{\isasymclose}\ %
\isamarkupcmt{comment in ML%
}{\isacartoucheclose}%
\endisatagML
{\isafoldML}%
\isadelimML
\endisadelimML
\begin{isamarkuptext}%
This approach of nesting languages resembles s-expressions in LISP, but the
  cartouche delimiters are visually less intrusive than parentheses, and the
  sub-expressions can be arbitrary sub-languages with their own concrete
  syntax. The Prover IDE helps users to understand complex nesting of
  languages, e.g. via text colors and popups (see \figref{fig:nesting}).
  Isabelle/ML helps language implementors with operations for common concepts,
  e.g. embedded comments seen here.

  \begin{figure}[htb]
  \centering
  \includegraphics[width=0.75\textwidth]{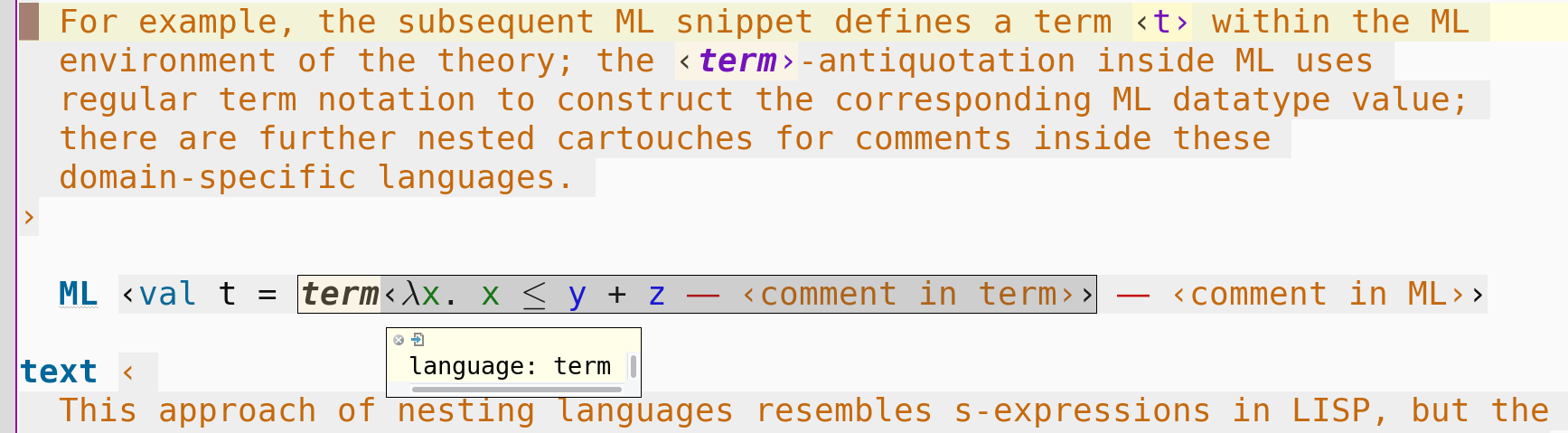}
  \caption{PIDE exploration of nested sub-languages within this paper}
  \label{fig:nesting}
  \end{figure}%
\end{isamarkuptext}\isamarkuptrue%
\isadelimdocument
\endisadelimdocument
\isatagdocument
\isamarkupsubsection{Auxiliary Files with Implicit Theory Context%
}
\isamarkuptrue%
\endisatagdocument
{\isafolddocument}%
\isadelimdocument
\endisadelimdocument
\begin{isamarkuptext}%
The outer syntax of Isabelle supports a special class of \emph{theory load}
  commands: there is a single argument that refers to a file, relative to the
  directory where the theory file is located. Isabelle/PIDE manages the
  content of that file in a stateless manner: the command implementation gets
  its source as attachment to outer syntax tokens --- there is no direct
  access to the file-system. In general, the editor could have unsaved buffers
  with changed content: the prover needs to process that intermediate state,
  not an old saved copy.

  Embedded source via auxiliary files is more scalable than inlined
  cartouches. For example, consider the commands \isa{\isacommand{ML}} vs. \isa{\isacommand{ML{\isacharunderscore}file}}: both
  incorporate Isabelle/ML definitions into the current theory context, and
  both are ubiquitous in the construction of Isabelle/Pure and Isabelle/HOL.
  \isa{\isacommand{ML}} is preferred for small snippets, up to one page of source text.
  \isa{\isacommand{ML{\isacharunderscore}file}} is better suited for big modules: Isabelle/jEdit provides a mode
  for the corresponding \isatt{.ML} files, with static syntax tables and a SideKick
  parser to generate tree views.

  Normally, theory load commands occur within a particular theory body to
  augment its content. In contrast, an \emph{implicit theory context} helps when
  the file is considered stand-alone: it refers to an imported context for its
  language definition, but the results of checking are restricted to PIDE
  markup shown to the user. The Isabelle/Scala programming interface allows to
  define a \emph{file-format} (according to the file extension): thus PIDE knows
  which theory template needs to be generated internally for such auxiliary
  files. Example file-formats are those for \isatt{bibtex} and \isatt{Naproche{\char`\-}SAD}
  (\secref{sec:naproche}).

  Still missing is support for simultaneous loading of files by a single
  command, e.g. a whole sub-project in an external language. That would
  require the Isabelle/Scala interface to understand the syntax of the load
  command, beyond a single file argument. Even more ambitious would be
  transitive exploration of included files, to refer to a complete graph via a
  few root entries in the text.%
\end{isamarkuptext}\isamarkuptrue%
\isadelimdocument
\endisadelimdocument
\isatagdocument
\isamarkupsubsection{Shallow Presentation of Document Sources%
}
\isamarkuptrue%
\endisatagdocument
{\isafolddocument}%
\isadelimdocument
\endisadelimdocument
\begin{isamarkuptext}%
PIDE interaction is about creating documents, and this is taken literally
for the final outcome: a traditional PDF produced via {\LaTeX}. An example
is the present paper itself: document sources are edited in
Isabelle/jEdit (see \figref{fig:nesting}) and the batch-mode tool
\isatt{isabelle\ build\ {\char`\-}o\ document=pdf} produces the typeset document for
publication. This works according to a rather shallow presentation scheme
going back to the early days of Isabelle/Isar (20 years ago), with a few
later additions. The idea is that the source language is sufficiently close
to a proper mathematical document, such that simple pretty-printing is
sufficient:

\begin{itemize}%
\item Isabelle symbols like \isatt{{\char`\\}{\char`\<}alpha{\char`\>}} are blindly replaced by {\LaTeX} macros: the
Isabelle style files provide a meaning for that to typeset \isa{{\isasymalpha}}.

\item Document \emph{markup commands} like ``\isa{\isacommand{section}\ {\isacartoucheopen}source{\isacartoucheclose}}'' or ``\isa{\isacommand{text}\ {\isacartoucheopen}source{\isacartoucheclose}}'' are turned into corresponding {\LaTeX} macros for sections,
paragraphs etc.

\item Document \emph{markdown items} with Isabelle symbols \isatt{{\char`\\}{\char`\<}{\char`\^}item{\char`\>}}, \isatt{{\char`\\}{\char`\<}{\char`\^}enum{\char`\>}}, \isatt{{\char`\\}{\char`\<}{\char`\^}descr{\char`\>}} are
turned into corresponding {\LaTeX} environments \isatt{itemize}, \isatt{enumerate},
\isatt{description}.

\item \emph{Embedded comments} like ``\isa{{\isasymcomment}}~\isa{{\isasymopen}source{\isasymclose}}'' are turned into suitable
{\LaTeX} macros from the Isabelle style files.

\item \emph{Document antiquotations} are evaluated and inlined: the user-defined
implementation in ML generates document output within the formal context,
e.g. to pretty-print a term using its logical notation.%
\end{itemize}

Note that genuine PIDE markup is not yet used for document output: it is
only available in Isabelle/Scala, but document preparation works in
Isabelle/ML. We can see that omission in the example of
\secref{sec:cartouches}: the typeset version of \figref{fig:nesting} does
not treat the ML keyword \isa{val} specifically in {\LaTeX}, it looks like a
regular identifier. Full semantic document preparation is an important area
of future work: presently there are only some experiments with HTML preview
in Isabelle/jEdit.

\smallskip PIDE editor presentation of document sources works differently: while the
prover is processing the sources, the online document model accumulates XML
markup over the original sources. This can be used for painting the editor
view in real-time, using whatever is available at a particular point in
time; formally this is a \emph{snapshot} of the PIDE document state. The jEdit
editor is a bit limited in its visual rendering capabilities, though: a
single font with small variations on style, and uniform font-size and
line-height. To make the best out of that, Isabelle/jEdit uses custom
Unicode fonts derived from the DejaVu collection, with mathematical symbols
taken from {\TeX} fonts. Isabelle2019 includes a standard set of font
families: \isatt{Sans\ Mono} (default for text buffers), \isatt{Sans} (default for GUI
elements), and \isatt{Serif} (default for help texts). To emphasize the
``ordinary mathematical text'' format of Naproche-SAD
(\secref{sec:naproche}), the screenshot in \figref{fig:isabelle-naproche}
has actually used the proportional \isatt{Sans} instead of the (almost) fixed
\isatt{Sans\ Mono}.

There are additional tricks in Isabelle/jEdit rendering to support
subscript, superscript, and bold-face of the subsequent Isabelle symbol, but
without nesting of font styles. Furthermore, there are some icons in the
font to render special control symbols nicely, e.g. \isatt{{\char`\\}{\char`\<}{\char`\^}item{\char`\>}} for Isabelle
Markdown as a square bullet, \isatt{{\char`\\}{\char`\<}{\char`\^}file{\char`\>}} as a sheet of paper, \isatt{{\char`\\}{\char`\<}{\char`\^}dir{\char`\>}} as a folder,
\isatt{{\char`\\}{\char`\<}{\char`\^}url{\char`\>}} as a W3C globe.

It is interesting to see how far the jEdit text editor can be stretched,
with the help of an open-ended collection of Isabelle symbols and
specifically generated application fonts. Compared to that, the modest
HTML5/CSS3 styling in Isabelle/VSCode is still lagging behind: the makers of
VSCode are taking away most of the rendering power of the underlying
Chromium browser, because they want to deliver an editor only for ``code'',
not documents.

\smallskip In the near future, there should be better convergence of offline PDF
presentation and online editor rendering of PIDE documents. In particular,
the Isabelle {\LaTeX} toolchain needs to be integrated into the IDE, to
avoid several seconds of wait time to produce PDFs. Further ideas for
renovation of Isabelle document preparation (mostly for HTML) are sketched
in \cite[\S3.3]{Wenzel:2018:Isabelle}.%
\end{isamarkuptext}\isamarkuptrue%
\isadelimdocument
\endisadelimdocument
\isatagdocument
\isamarkupsection{Aims and Approaches of Isabelle/PIDE \label{sec:approaches}%
}
\isamarkuptrue%
\endisatagdocument
{\isafolddocument}%
\isadelimdocument
\endisadelimdocument
\begin{isamarkuptext}%
What has Isabelle/PIDE tried to achieve over the past 10 years, what worked
  out and what failed? The subsequent overview of important aims and
  approaches summarizes \textbf{success}, \textbf{failure}, \textbf{changes} in the plan, and
  ideas for \textbf{future} work.%
\end{isamarkuptext}\isamarkuptrue%
\isadelimdocument
\endisadelimdocument
\isatagdocument
\isamarkupsubsection{Isabelle/ML vs. Isabelle/Scala: ``Mathematics'' vs. ``Physics''%
}
\isamarkuptrue%
\endisatagdocument
{\isafolddocument}%
\isadelimdocument
\endisadelimdocument
\begin{isamarkuptext}%
At the bottom, Isabelle is an LCF-style proof assistant \cite{Gordon-Milner-Wadsworth:1979} that is freely programmable in Isabelle/ML.
  That is based on Poly/ML, which is well-tuned towards applications of
  symbolic logic on multicore hardware.

  To complement the ultra-pure ML environment by tools and libraries from the
  ``real'' world, Isabelle/PIDE has been based on Scala/JVM from the very
  beginning in 2008. The JVM gives access to GUI frameworks, HTTP servers,
  database engines (SQLite, PostgreSQL) etc. The programming style of
  Isabelle/Scala follows that of Isabelle/ML to a large extent, and there are
  many basic libraries that are provided on both sides.

  {\success} The clean and efficient functional style of ML has been
  transferred to Scala. There are many modules on both sides that follow the
  typical Isabelle mindset of minimality and purity. It is feasible to move
  the language boundary of tool implementations, according to technical
  side-conditions of ML vs. Scala.

  {\failure} Isabelle users often find Isabelle/ML as tool implementation
  language already too difficult. The additional Isabelle/Scala for tool
  integration is beyond the multilingual capabilities of most people. This
  could be partly caused by common misunderstandings about both sides:
  Isabelle/ML is not just Standard ML, and Isabelle/Scala not just Scala ---
  both are ``Isabelle'' with an idiomatic style that deviates from customs
  seen elsewhere.

  {\changes} The original conception of Isabelle/Scala as add-on library for
  system integration turned out insufficient. Instead, Isabelle/Scala and
  Isabelle/ML have become equal partners in forming the Isabelle
  infrastructure. Consequently, Isabelle/Pure now contains many ML and Scala
  modules side-by-side, sometimes with equivalent functionality (e.g. portable
  file and process operations), and sometimes complementary (e.g. for the PIDE
  protocol).

  {\future} Isabelle/Scala still needs proper IDE integration: its development
  model resembles that of Isabelle/ML in earlier decades. It is a funny
  paradox that the Prover IDE infrastructure is developed with a plain text
  editor and command-line build process. Either Scala could be integrated into
  PIDE as another back-end, or a regular Scala IDE could be used (e.g.
  IntelliJ IDEA).%
\end{isamarkuptext}\isamarkuptrue%
\isadelimdocument
\endisadelimdocument
\isatagdocument
\isamarkupsubsection{Private PIDE protocol (untyped) vs. public APIs (typed)%
}
\isamarkuptrue%
\endisatagdocument
{\isafolddocument}%
\isadelimdocument
\endisadelimdocument
\begin{isamarkuptext}%
Isabelle/PIDE resides both in Isabelle/ML and Isabelle/Scala, with typed
  functional programming interfaces. The PIDE implementation uses a
  custom-made protocol that fits tightly to the requirements of the
  interactive document model. Over the years, there have been frequent changes
  and adjustments of the protocol. The communication works over a pure
  byte-channel, with low overhead for structured messages and ML-like datatype
  values. The paper \cite{Wenzel:2013:CoqPIDE} explains the PIDE protocol
  for demonstration purposes with a back-end in Coq 8.4 (2013).

  {\success} Efficient and robust implementation of the bi-lingual PIDE
  framework in Isabelle/ML/Scala works. Easy maintenance of corresponding
  modules in the same directory location is feasible.

  {\failure} It is cumbersome to develop and maintain different PIDE
  implementations for different provers: the Coq/PIDE project \cite{Tankink:2014:UITP-EPTCS} did not reach end-users and is now lagging behind
  years of further PIDE development.

  {\changes} The initial conception of the PIDE protocol was quite basic, but
  it acquired complexity and sophistication over time.

  {\future} Back-end protocol: the old idea to retarget PIDE for other provers
  (like Coq) could be re-opened eventually, but it requires significant
  personal dedication and resources to do that properly. Front-end protocol:
  there is already a simplified public PIDE protocol for headless interaction
  with the document-model. That could eventually become a client-server
  protocol for web applications.%
\end{isamarkuptext}\isamarkuptrue%
\isadelimdocument
\endisadelimdocument
\isatagdocument
\isamarkupsubsection{Pervasive Parallelism on Multicore Hardware%
}
\isamarkuptrue%
\endisatagdocument
{\isafolddocument}%
\isadelimdocument
\endisadelimdocument
\begin{isamarkuptext}%
Both Isabelle/ML (``pure mathematics'') and Isabelle/Scala (``real
  physics'') support parallel programming with shared memory and immutable
  values. User-space tools do not have to care much about it, as long as
  standard Isabelle programming idioms and libraries are used. Scaling of
  parallel programs is always a challenge: Isabelle/ML performs well into the
  range of 8--16 cores (on a single CPU node). Isabelle/Scala rarely uses more
  than 2--4 cores.

  {\success} Parallel Isabelle/ML became routinely available in 2008 \cite{Wenzel:2009}, and has been refined many times \cite{Matthews-Wenzel:2010,Wenzel:2013:ITP}. That proved so successful
  that an initial motivation for PIDE was to make an IDE that can properly
  connect to a parallel proof engine: for the user front-end this added the
  aspect of asynchronous interaction, which is central to PIDE \cite{Paral-ITP:CICM2013,Wenzel:2014:ITP-PIDE}.

  {\failure} The predictions of CPU manufacturers in 2005 about consumer
  machines with many cores (32--128) have not become true, because mainstream
  applications cannot use so much parallelism. Instead we have seen a trend
  towards light-weight mobile devices (with 2--8 cores). This can confuse new
  Isabelle users: they think that big applications from AFP should work on
  e.g. 2 CPU cores and 4 GB RAM, but reality demands to double or quadruple
  these resources. When we see server-class machines with a lot of cores,
  there is often an internal division into separate CPU nodes (NUMA) with
  significant penalty for a shared-memory application, e.g. a machine with 64
  cores might turn out as 8 nodes of 8 core CPUs with delay factor 1.6--3.2 to
  access data on a distant node.

  {\future} The Isabelle/PIDE front-end (on a small mobile device) and the
  back-end (on a big server) could be separated, to follow a general trend
  towards ``cloud computing''. This could be done without degrading the IDE
  into a mere web-browser application. Instead, the existing Isabelle/jEdit or
  Isabelle/VSCode desktop applications could connect to a remote version of
  Isabelle/PIDE, see also \cite[\S3.3]{Wenzel:2018:Isabelle}. Headless
  PIDE functionality has already been implemented and used elsewhere, e.g. to
  export formal content of AFP entries in Isabelle/MMT \cite[\S3.1]{CICM-2019:RDF}. A proper client-server environment across the Web would
  require substantial work, e.g. robust management of lost connections to the
  remote back-end.%
\end{isamarkuptext}\isamarkuptrue%
\isadelimdocument
\endisadelimdocument
\isatagdocument
\isamarkupsubsection{Desktop Application Bundles on Linux, Windows, macOS%
}
\isamarkuptrue%
\endisatagdocument
{\isafolddocument}%
\isadelimdocument
\endisadelimdocument
\begin{isamarkuptext}%
Isabelle/PIDE is available as a single download that can be unpacked and run
  without further ado. There is no requirement for self-assembly by end-users
  Isabelle does not need packaging by OS providers (e.g. Debian): it is a
  genuine end-user application; there can be different versions side-by-side
  without conflict.

  {\success} The majority of users is happy with the all-inclusive Isabelle
  application bundle (2019: 300\,MB download size, 1\,GB unpacked size). It
  works almost as smoothly as major Open Source products (e.g. Firefox,
  LibreOffice).

  {\failure} Full equality of all three platforms families has still not been
  achieved. Java GUI rendering on Linux is worse than on Windows and macOS;
  exotic Linux/X11 window managers can cause problems. Add-on tools are
  sometimes not as portable (and robust) on Windows: Isabelle often refers to
  Cygwin as auxiliary POSIX platform, but that may cause its own problems.
  Even for native Windows tools (e.g. via MinGW) the Unix-style orchestration
  of multiple processes can have timing problems (e.g. Sledgehammer provers)
  or fail due to antivirus software. Moreover, the recent tendency of Windows
  and macOS towards ``application stores'' makes it harder to run downloaded
  Isabelle bundles on the spot: users need to bypass extra vendor checks.

  {\changes} The initial approach was more optimistic about availability of
  certain ``standard'' components on the OS platform, notably Java and Scala.
  Later it became clear that almost everything needs to be bundled with
  Isabelle, except for the most basic system components (e.g. \isatt{libc},
  \isatt{libc++}, \isatt{curl}, \isatt{perl}). Note that other projects have come to a
  similar conclusion, e.g. SageMath with very thorough all-inclusive bundling.

  {\future} The ``download--unpack--run'' experience of Isabelle/PIDE needs
  fine-tuning for first-time users. In particular, AFP needs to be included in
  this, to avoid manual intervention with session \isatt{ROOTS} and \isatt{ROOT} files.
  There could be more support for alternative applications based on the
  Isabelle/PIDE platform, to suppress unused components of Isabelle/HOL, e.g.
  see Isabelle/Naproche (\secref{sec:naproche}).%
\end{isamarkuptext}\isamarkuptrue%
\isadelimdocument
\endisadelimdocument
\isatagdocument
\isamarkupsection{Conclusion%
}
\isamarkuptrue%
\endisatagdocument
{\isafolddocument}%
\isadelimdocument
\endisadelimdocument
\begin{isamarkuptext}%
Isabelle/PIDE is a long-term effort to support live editing of complex
document structures with ``active'' content. Its cultural background is that
of interactive theorem proving, which has high demands for execution
management: interrupts, real-time requirements, parallel threads, external
processes. So this provides a generous upper bound of technology for less
ambitious applications of PIDE. We have seen the example of
Isabelle/Naproche.

Ultimately, we may understand PIDE as a continuation and elaboration of the
following approaches:

\begin{itemize}%
\item The LCF/ML approach to interactive theorem proving by Milner et-al
\cite{Gordon-Milner-Wadsworth:1979,Gordon-Milner-etal:1978}.

\item The Isar approach to human-readable proof documents by Wenzel \cite{Wenzel:2006:Festschrift}.

\item Parallel ML and future proofs by Matthews and Wenzel \cite{Wenzel:2009,Matthews-Wenzel:2010,Wenzel:2013:ITP}.

\item Early prover interfaces by Aspinall \cite{Aspinall-et-al:2007},
Bertot \cite{Bertot-Thery:1996} and others.%
\end{itemize}

All of that taken together amounts to decades of research: PIDE attempts to
form a limit over that, to reach a new stage of semantic editing that can be
taken for granted. The present paper has illustrated some applications and
sketched the overall construction. It is to be hoped that builders of other
mathematical tools are encouraged to re-use PIDE for their own projects.%
\end{isamarkuptext}\isamarkuptrue%
\isadelimtheory
\endisadelimtheory
\isatagtheory
\isacommand{end}\isamarkupfalse%
\endisatagtheory
{\isafoldtheory}%
\isadelimtheory
\endisadelimtheory
\end{isabellebody}%

\bibliographystyle{splncs04}
\bibliography{root}

\end{document}